\documentclass[notitlepage,letterpaper,final]{iopart}
\usepackage{iopams}

\usepackage{graphicx}
\usepackage{color}
\usepackage[utf8]{inputenc}
\usepackage[pdftex,bookmarks=true,backref=page,pagebackref=true,colorlinks=true, linkcolor=blue]{hyperref}

\newcommand{\code}{M3D-$C^1$}
\newcommand{\text}{\textrm}

\begin{document}

\title{Modeling of carbon pellets disruption mitigation in an NSTX-U plasma}

\author{C F Clauser$^{1,2}$, S C Jardin$^2$, R Raman$^3$, B C Lyons$^4$, N M Ferraro$^2$}
\address{$^1$Lehigh University, Bethlehem, Pennsylvania 18015, USA}
\address{$^2$Princeton Plasma Physics Laboratory, Princeton, New Jersey 08543, USA}
\address{$^3$University of Washington, Seattle, Washington 98195, USA}
\address{$^4$General Atomics, San Diego, California 92121, USA}
\ead{cclauser@lehigh.edu}

\begin{abstract}
Single carbon pellet disruption mitigation simulations using \code were conducted in an NSTX-U-like plasma to support the electromagnetic pellet injection concept (EPI). A carbon ablation model has been implemented in \code and tested with available data. 2D simulations were conducted in order to estimate the amount of carbon needed to quench the plasma, finding that the content in a $1\,$mm radius vitreous carbon pellet ($\sim 3.2\times 10^{20}\,$atoms) would be enough if it is entirely ablated. 3D simulations were performed, scanning over pellet velocity and parallel thermal conductivity, as well as different injection directions and pellet concepts (solid pellets and shell pellets). The sensitivity of the thermal quench and other related quantities to these parameters has been evaluated. A $1\,$mm radius solid pellet only partially ablates at velocities of $300\,$m/s or higher, thus being unable to fully quench the plasma. To further enhance the ablation, approximations to an array of pellets and the shell pellet concept were also explored. 3D field line stochastization plays an important role in both quenching the center of the plasma and in heat flux losses, thus lowering the amount of carbon needed to mitigate the plasma when compared to the 2D case. This study constitutes an important step forward in `predict-first' simulations for disruption mitigation in NSTX-U and other devices, such as ITER.
\end{abstract}


\vspace{2pc}\noindent\textit{ Keywords\/}: 
Tokamak,
Disruption Mitigation,
Pellet injection,
Extended MHD, 
M3D-$C^1$
\newline


\ioptwocol
\section{Introduction}
\label{sec:intro}

It is well established that ITER and future tokamaks must be equipped with an effective method for injecting impurities for rapid shutdown and to mitigate the damage caused by plasma disruptions \cite{Hender2007}.  The present baseline concept for this injection system on ITER is shattered pellet injection (SPI) \cite{Commaux2016}.

As an alternative to shattered pellet injection, an electromagnetic pellet injection (EPI) device has been recently proposed \cite{Raman2019,Raman2021}. The EPI system accelerates a sabot electromagnetically with a rail gun. The sabot is a metal capsule that can be accelerated to high-velocities.   At the end of its acceleration, the sabot will release the radiative payload that is composed of granules of low-Z materials, or a shell pellet containing smaller pellets.  EPI would offer a fast response time and high enough speed to deposit the payloads in the plasma core in ITER.

To understand the physics involved, reliable simulations that can evaluate and predict the evolving plasma in this situation are essential. Recently, the \code code has incorporated impurity radiation and pellet injection modules \cite{Lyons2019,Ferraro2019} which allows it to perform these kinds of studies, and benchmark exercises are presently underway.

To explore the EPI concept and its potential benefits, we have conducted a series of simulations modeling the injection of a single carbon pellet in NSTX-U. To do this, a carbon ablation model \cite{Sergeev2006} was incorporated in \code. As a first step, the ablation model was tested by performing a simulation of carbon injection into an ASDEX-U discharge  for which data exists \cite{Sergeev2006}. Next, we performed a convergence study for NSTX-U covering different modeling parameters. We compare these cases and show the sensitivity to the induced thermal quench and other relevant parameters on the physical input and modeling parameters.

This paper is organized as follows: in Sec. \ref{Sec:pellet_modelling} we briefly summarize the modeling of pellets in \code and how it is coupled to the MHD equations. We also describe the carbon model that was implemented in \code. In Sec. \ref{Sec:validation} we present a comparison of our implementation with available data from ASDEX-U. Section \ref{Sec:NSTX-U} presents the NSTX-U equilibrium configuration used in the remaining studies.   Section \ref{sec:NSTX_2D} presents some preliminary 2D modeling results for subsequent comparisons.  In Sec. \ref{sec:NSTX_3D} we present C-pellet disruption mitigation simulations for proposed NSTX-U experiments and show some sensitivities to modeling and physical parameters. Finally, we summarize the results in Sec. \ref{Sec:conclusions}.

\section{Pellet-injection modeling in \code}
\label{Sec:pellet_modelling}

For convenience, we summarize here the physics and modeling involved in \code, focusing on disruption mitigation studies.  These have been extensively described in Refs. \cite{Ferraro2019,Lyons2019}. \code is a non-linear 3D extended-MHD code that uses high-order finite elements and implicit time-stepping to advance the equations in time. \cite{Jardin2012}. 

 The physical quantity that \code uses to incorporate the physics of pellet injection is the ablation rate, $\dot{N}$, which has to be provided, usually as a function of the plasma temperature and density in the vicinity of the pellet position. 
 
The ablated material is weighted with a prescribed spatial distribution in the region surrounding the pellet position and is included as a source in the fluid equations. Different spatial distribution functions have been incorporated in \code. In this work, we have used a Gaussian-like shape distribution so that the source term for the ablated material reads
\begin{equation} 
\eqalign{
\sigma = \frac{\dot{N}}{(2\pi)^{3/2}\Delta_{p}^{2}\Delta_t} 
                \exp \bigg\{ -\frac{(R-R_p)^2+(Z-Z_p)^2}{2\Delta_{p}^{2}}  \cr
                \qquad \qquad \qquad  \qquad \quad
                    -\frac{RR_p(1-\cos(\varphi - \varphi_p)}{\Delta_{t}^{2}} \bigg\}
} \label{Eq:ablated_distribution}
\end{equation}
where cylindrical coordinates $(R,\varphi,Z)$ are employed and the (time dependent) pellet position is given by $(R_p,\varphi_p,Z_p)$. 
The typical size of the ablated material cloud is specified by $\Delta_p$ (poloidal width) and $\Delta_t$ (toroidal width). Limitations on these values arise due to the mesh resolution that \code uses to solve the fluid equations in 3D. \\
The ablated material is initialized as neutrals and all the atomic physics (ionization, recombination, radiation, etc) is calculated with the KPRAD module that is coupled to the extended MHD equations \cite{Ferraro2019,Lyons2019}. Thus, different source terms $\sigma_s$ are generated for main ions and each charge state of impurities (here, each charge state is considered a different species). \\
These source terms and the radiated energy are coupled to the fluid equations to evolve the plasma in a self-consistent way. 
The implementation solves a continuity equation for each charge state of each ionized species (main ions and impurities),
\begin{equation}
\frac{\partial n_s}{\partial t} + \nabla \cdot (n_s {\bf v}) 
                     =   \nabla \cdot D \nabla n_s  + \sigma_s .
\end{equation}
Here, $n_s$ is the density of each charge state of each plasma species $s$, $D$ is a density diffusion coefficient usually employed for numerical stability, $\sigma_s$ is the source term calculated with the KPRAD module for each charge state. The electron density is defined to satisfy the quasi-neutrality condition. \\
The coupling with the momentum equation was implemented in a single-fluid velocity model, i.e. all species have the same fluid velocity ${\bf v}$,
\begin{equation}
\rho \left( \frac{\partial {\bf v}}{\partial t} + {\bf v}\cdot \nabla {\bf v} \right) 
                     =  {\bf J} \times {\bf B} - \nabla p - \nabla \cdot \Pi - \bar{\omega}{\bf v}
\end{equation}
where
\begin{eqnarray*}
\rho = m_i n_i + \sum_{j=1}^{Z} m_z n_{z}^{(j)}  \\
\bar{\omega} = m_i \sigma_i + \sum_{j=1}^{Z} m_z \sigma_{z}^{(j)} .
\end{eqnarray*} 
Here $j$ represents the charge state of an impurity with nuclear charge $Z$. The quantity $\bar{\omega}{\bf v}$ appears as a consequence of momentum conservation in the single fluid approximation.\\
The momentum equation for electrons leads to the Ohm's law
\begin{equation}
{\bf E} = \eta {\bf J} - {\bf v}\times {\bf B}.
\end{equation}

For the energy equations, there are different assumptions that were implemented, as explained in Ref. \cite{Ferraro2019}.
Here, we used the two temperature model, which evolves a temperature for the electrons:
\begin{equation}
\eqalign{
\fl n_e \left( \frac{\partial T_e}{\partial t} + {\bf v}\cdot \nabla T_e + (\Gamma - 1)T_e \nabla \cdot {\bf v} \right) +  \cr
 \quad           T_e ( \nabla \cdot D \nabla n_e + \sigma_e) =  \cr
 \qquad           (\Gamma - 1) \left(\eta {\bf J}^2 - \nabla \cdot {\bf q}_e + Q_e + Q_{\Delta} - \Pi_e : \nabla {\bf v} \right),
 }
\end{equation}
and also a temperature for all the ion species:
\begin{equation}
\eqalign{
\fl n_* \left( \frac{\partial T_i}{\partial t} + {\bf v}\cdot \nabla T_i + (\Gamma - 1)T_i \nabla \cdot {\bf v} \right) +  \cr
 \quad           T_i ( \nabla \cdot D \nabla n_* + \sigma_*) =  \cr
 \qquad           (\Gamma - 1) \bigg(  - \nabla \cdot {\bf q}_* + Q_* - Q_{\Delta} - \cr
\qquad \qquad \qquad \qquad \qquad  \left.  \Pi_* : \nabla {\bf v} + \frac{1}{2}\bar{\omega}v^2 \right).
}
\end{equation}
Here $n_*$, $\sigma_*$, $\Pi_*$, $Q_*$ and ${\bf q}_*$ are sum over all ion species of the particle densities, particle source densities, stresses, energy density sources, and energy density fluxes, respectively. All the ions (main plasma ions and ionized impurities) are assumed to have the same temperature.
For a detailed explanation of these equations, see Ref. \cite{Ferraro2019}

Several ablation models have been derived for different pellet materials \cite{Parks1977,Parks1978,Sergeev2006} in order to provide the quantity $\dot{N}$.
Many of these rely on the neutral gas shielding approximation and scaling laws.
For carbon pellets we implemented the model described in Refs. \cite{Kuteev1984,Sergeev2006}. This model is based on the so-called shielding factor, which is the ratio of the plasma heat flux that enters into the neutral cloud and the heat flux that actually reaches the pellet surface. In Ref. \cite{Sergeev2006}, Sergeev \etal present ablation rates for both strong and weak shielding limits and they propose a simple interpolation formula for intermediate shielding to cover both limiting cases (see Eq. (25) in the same reference). As a first step towards carbon ablation simulations, we have implemented this ablation model in \code.
Recently, other ablation models have been used for carbon granule injection (and other materials) \cite{Lunsford2019} and could be also implemented in \code.

\section{Validating the ablation rate implementation in \code}
\label{Sec:validation}
The ablation model described in Ref. \cite{Sergeev2006} has been used to analyze several discharges \cite{Kuteev1984,Sergeev1994,Sergeev2006}. In order to test the implementation in \code, we have chosen one of those cases: an ASDEX--U-like plasma, which was reconstructed from the shot $\#3948$ using the available information \cite{Sergeev2006}.
\begin{figure}[h!]
\centering
\includegraphics[width=0.9\columnwidth]{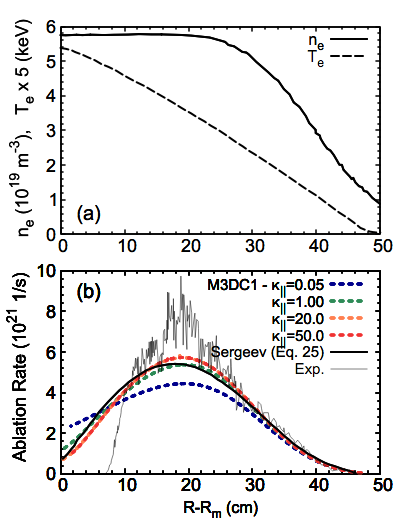}
\caption{Comparison of the ablation rate obtained in \code in an ASDEX-U-like plasma for which data existed. In (a) the plasma kinetic profiles are shown and figure (b) shows the ablation rate obtained with \code for different parallel heat flux coefficients $\kappa_\parallel$, compared with results reported in Ref. \cite{Sergeev2006}.  }
\label{fig:AUG}
\end{figure}

The pellet had an initial radius $r_p = 0.25\,$mm and was injected from the outer mid-plane region towards the plasma core. In our simulations we approximated it as a radially inward injection from the outer mid-plane. The pellet velocity was $v_p = 485\,$m/s. 
Fig. \ref{fig:AUG} shows (a) the initial kinetic profiles and (b) the ablation rate, both as a function of the distance to the magnetic axis. Since the pellet is small, no significant perturbations to the plasma kinetic profiles were observed. 
The thin black curve in Fig. \ref{fig:AUG}(b) is the experimental signal and the thicker black curve is the result reported in \cite{Sergeev2006}. The four colored dashed curves are the \code results, in which different parallel heat flux coefficients, $\kappa_{\parallel}$, were used. We observe that the agreement is very good if the parallel heat flux is large enough, comparable to that in the experiment [in these simulations we used $\kappa_\perp = 3\times 10^{-6}$. To obtain $\kappa$ values is SI units, multiply by $1.54\times 10^{26}\,\text{m}^{-1}\text{s}^{-1}$]. \\
We note here that, for very low parallel heat flux, the ablation rate becomes smaller because the heat flux is not large enough to keep the temperature surrounding the pellet high. The local temperature around the pellet becomes colder due to the radiation of the ablated material. When increasing the parallel heat flux coefficient, the ablation increases because the plasma can quickly offset the radiated energy due to the ablated material, increasing the local temperature. At some point, increasing the parallel heat flux coefficient even more will not produce an increase in the ablation rate since the local temperature is completely balanced by the parallel heat flux and the temperature becomes uniform in a flux surface.
However, as will be shown below, stochastic field lines caused by the pellet transit can change this behavior. 

\section{ NSTX-U Equilibrium}
\label{Sec:NSTX-U}

\begin{figure}[h!]
\centering
\includegraphics[width=0.9\columnwidth]{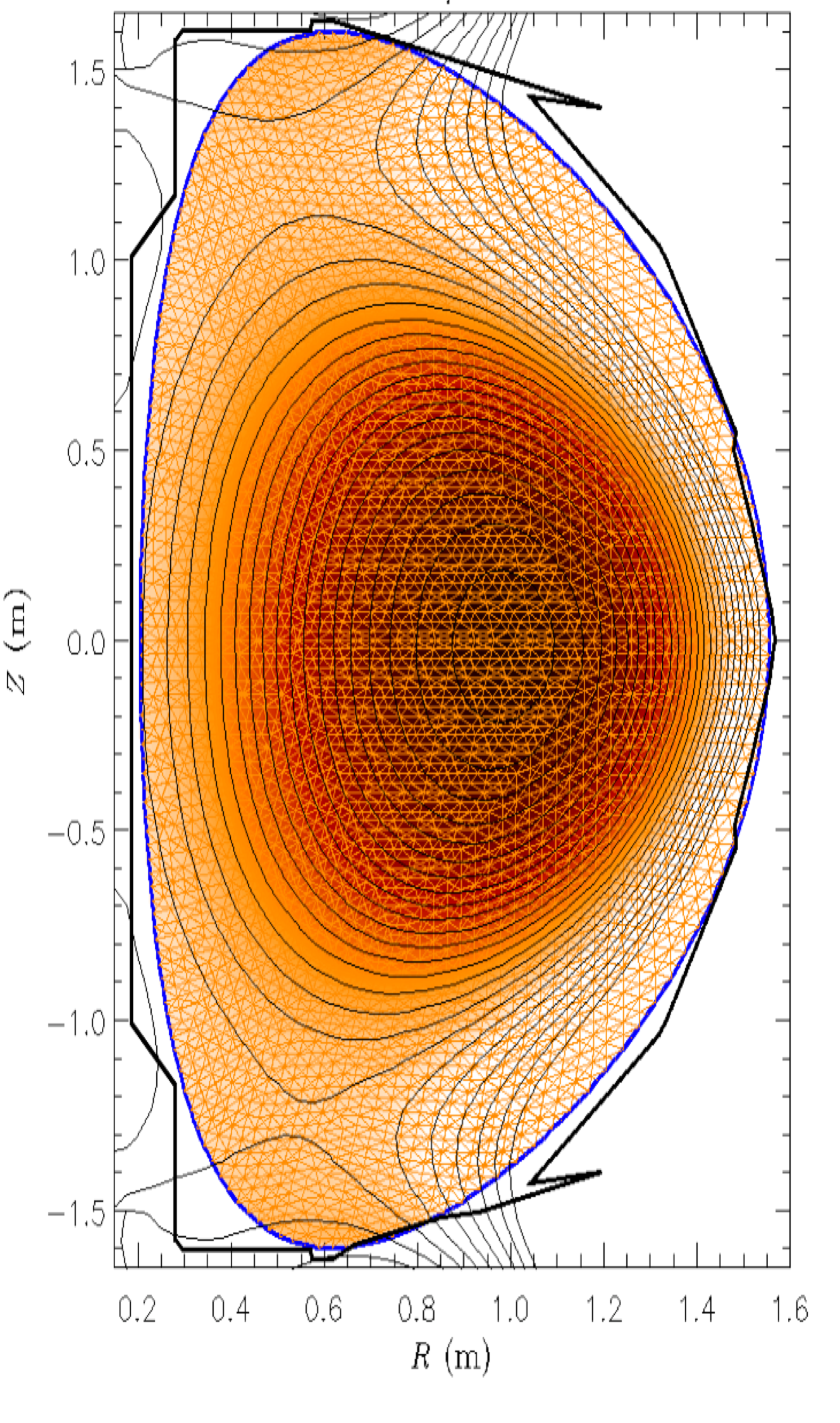}
\caption{Computational domain, mesh and initial equilibrium configuration for the NSTX-U geometry employed. Ideal boundary conditions were used.}
\label{Fig:mesh}
\end{figure}

The good agreement presented in the previous section provides validation to conduct a series of a ``predict-first'' simulations in which a single carbon pellet is injected into a NSTX-U-like discharge to instigate a thermal quench. We scanned over pellet injection conditions that support the EPI concept.
Figure \ref{Fig:mesh} shows the simulation domain, including the meshed plasma region in orange and boundary contour in blue. In addition, poloidal flux equilibrium contours are shown together with the actual NSTX-U wall in black as a reference.
For these simulations we have used an ideal wall boundary condition which, as can be seen from the figure, approximates the NSTX-U wall. The equilibrium has a magnetic major radius $R_m = 0.99\,$m, vacuum magnetic field at $R_m$ of $B_0 = 0.44\,$T, plasma current $I_p = 580\,$kA and $\beta = 2.25\,\%$. The total initial thermal energy is $69\,$kJ and the plasma magnetic energy is $120 \,$kJ.

\begin{figure}[h!]
\centering
\includegraphics[width=0.9\columnwidth]{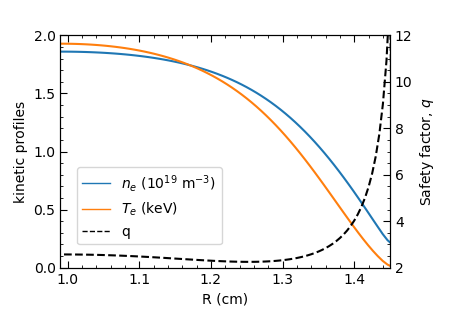}
\caption{Initial kintetic profiles and safety factor at the midplane, as a function of the mayor radius. Magnetic axis is at $R=0.99\,$m.}
\label{Fig:nstxu_kin_prof}
\end{figure}

Figure \ref{Fig:nstxu_kin_prof} shows the initial kinetic profiles, temperature and density, as well as the initial safety factor profile as a function of the major radius.
The boundary temperature was set to $1\,$eV. 

\section{NSTX-U Disruption Mitigation: 2D preliminary studies}\label{sec:NSTX_2D}

As a first step towards fully 3D carbon pellet injection simulations, we conducted a series of 2D simulations without a pellet but with an initial distribution of carbon atoms. 
This was performed in order to have a proxy of the amount of carbon that is needed to mitigate the plasma.
\begin{figure}[h!]
\centering
\includegraphics[width=0.9\columnwidth]{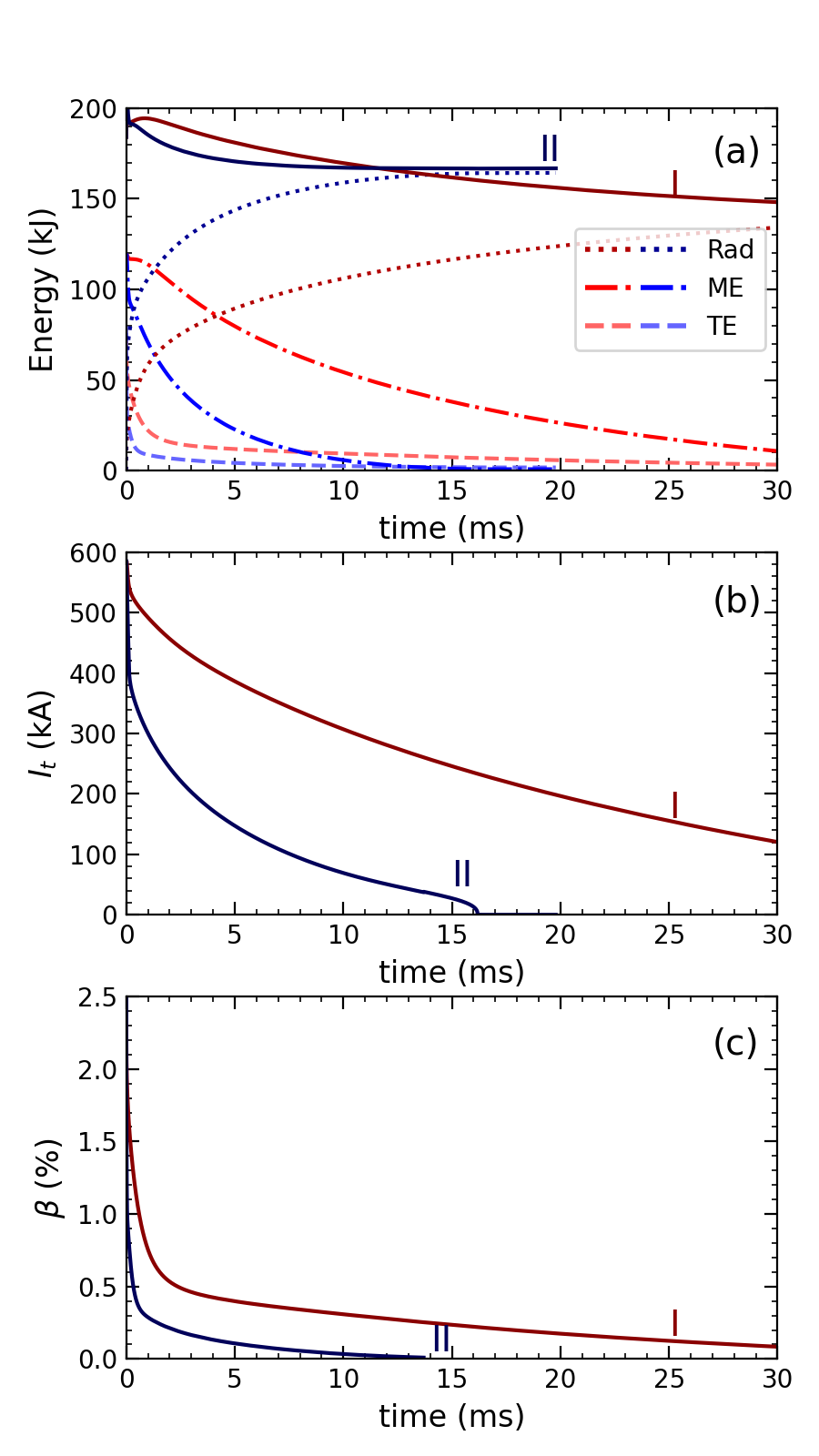}
\caption{2D simulations with an initial distribution of carbon equals to (I) the initial electron density and (II) twice the initial electron density. (a) shows the thermal energy (TE), the magnetic energy (ME) and the radiated energy (Rad). The solid lines are the sum of the three of them. (b) shows the plasma toroidal current, and (c) the plasma $\beta$.}
\label{Fig:2D_global}
\end{figure}
The initial distribution of carbon atoms was set to be proportional to the electron density, i.e. with the same spatial distribution but scaled with a constant. Figure \ref{Fig:2D_global} shows the time evolution of global quantities for two cases: with an initial carbon density equals to (I) the initial electron density and (II) twice the initial electron density. The amount of carbon in case (I) is  $2\times 10^{20}\,$atoms while in case (II) is $4\times 10^{20}\,$atoms. As a reference, a $1\,$mm-radius vitreous-carbon pellet ($\rho=1.51\,$g/cm$^{3}$) has $3.2\times 10^{20}$ atoms, which is between these two cases.\\
Figure \ref{Fig:2D_global}(a) shows different components of the total energy for each case: the plasma thermal energy (TE) in dashed lines, plasma magnetic energy (ME) in dash-dotted lines, the radiated energy (Rad) in dotted lines, and the addition of all of these components (solid lines). The drop in this solid lines represent the heat that went into the wall (heat flux). For case (II) this was only $\sim 25\,$kJ or $13\%$ of the total plasma energy. Figures \ref{Fig:2D_global}(b) and (c) show the plasma current and plasma $\beta$, respectively. We observe that in both cases the plasma is quenched (particularly case (II)), suggesting that the amount of carbon in $1\,$mm-radius pellet would be enough to mitigate the plasma if it were entirely ablated.

\section{NSTX-U Disruption Mitigation via pellet injection}
\label{sec:NSTX_3D}

\begin{figure}[h!]
\centering
\includegraphics[width=0.9\columnwidth]{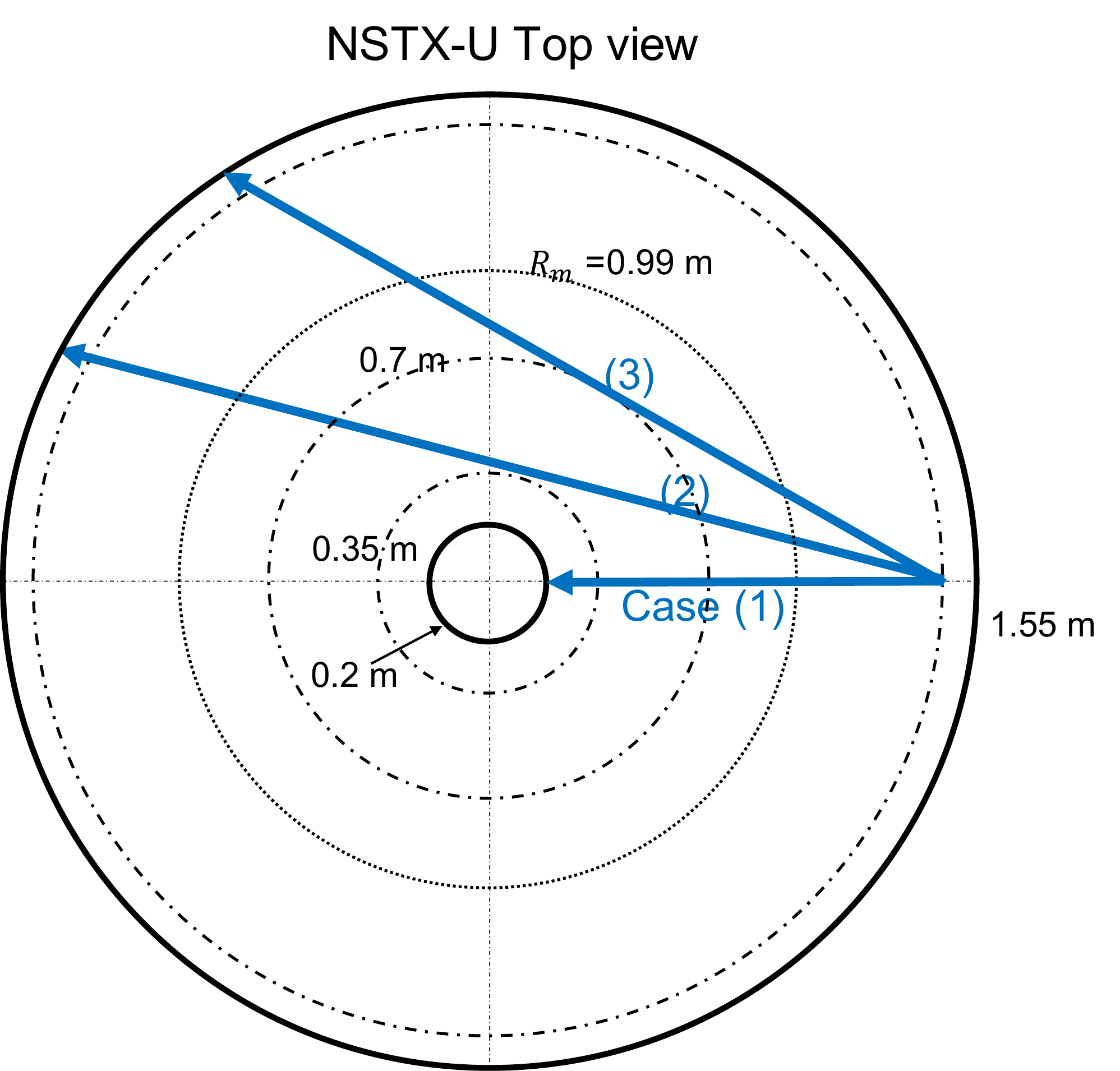}
\caption{Schematic NSTX-U top view showing the different injection directions chosen in the simulations. Case (1 - 3) are referred in Secs. \ref{sec:scan_velocity} -  \ref{sec:shell_pellet}, respectively.}
\label{Fig:top_view}
\end{figure}

Based on the previous 2D simulations, we have taken this amount of carbon in a $1\,$mm-radius pellet ($3.2\times 10^{20}\,$atoms) as a reference case in this study. In this section we present several cases of carbon pellet injection scanning over different parameters. The injection was from the outer mid-plane. Figure \ref{Fig:top_view} shows a schematic top-view of the device. Blue arrows represent the injection directions considered in this work. Case (1) will be discussed in Sec. \ref{sec:scan_velocity} and \ref{sec:scan_kappa}, while Case (2) will be discussed in \ref{sec:pellet_array}.  Case (3)
is a shell pellet which will be discussed in Sec. \ref{sec:shell_pellet}

Regarding the ablated material cloud size (see Eq. \eref{Eq:ablated_distribution}) , we have chosen $\Delta_t = 50\,$cm and $\Delta_p = 5\,$cm. Even though it is not shown here, reducing the ablated cloud even more does not lead to a significant change in global quantities but requires much smaller time steps and increased spatial resolution making the simulations much more expensive. The density diffusion term in the continuity equation for each species ranged between $2-10 \times 10^{-5}$ (internal units) to avoid numerical instabilities. The plasma viscosity was taken to be $5\times10^{-5}$ (internal units). These quantities can be reduced using a finer mesh but that would require much more computational resources, being unpractical for convergence studies. Further studies targeted to a particular configuration might have smaller values in these modelling parameters.
\begin{figure*}[h!]
\centering
\includegraphics[width=1.9\columnwidth]{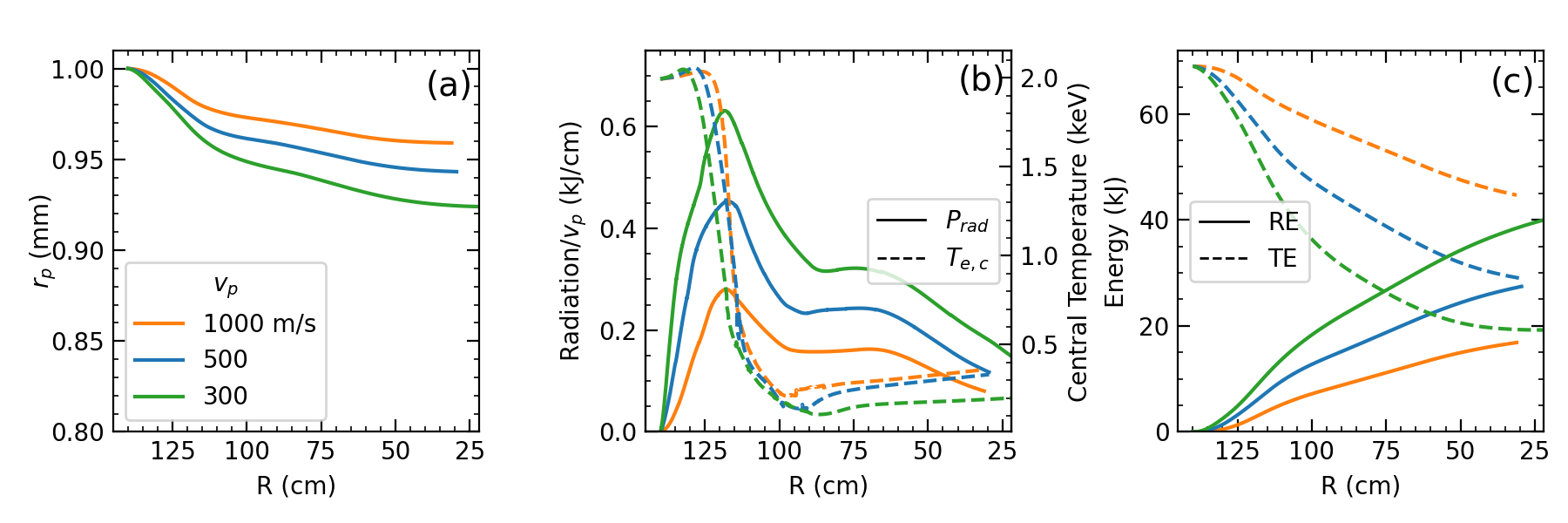}
\caption{Scanning over three pellet velocities: $1000$, $500$ and $300\,$m/s, as a function of the pellet position. (a) pellet radius, (b) radiated power normalized by the pellet velocity (solid lines) and plasma central electron temperature (dashed lines), (c) radiated energy (RE) and plasma thermal energy (TE).}
\label{Fig:global_vp}
\end{figure*}

\subsection{Case 1: Scan over pellet velocity}\label{sec:scan_velocity}
Figure \ref{Fig:global_vp} shows different global quantities as a function of the pellet position for three different velocities: $1000$, $500$ and $300\,$m/s. In all these cases, $\kappa_{\parallel}$ was set to $1$. As a reference, the outer wall is at approximately  $R=1.55\,$m while the inner wall is at $R=0.2\,$m. The initial pellet position was set to $R=1.4\,$m which is in the inner side of the separatrix. Figure \ref{Fig:global_vp}(a) shows the pellet radius. Fig. \ref{Fig:global_vp}(b) shows the radiated power normalized by the pellet velocity (solid lines) and the plasma electron central temperature (dashed lines). Fig. \ref{Fig:global_vp}(c) shows the thermal (TE) and radiated (RE) energy.

\begin{figure}[t!]
\centering
\includegraphics[width=0.85\columnwidth]{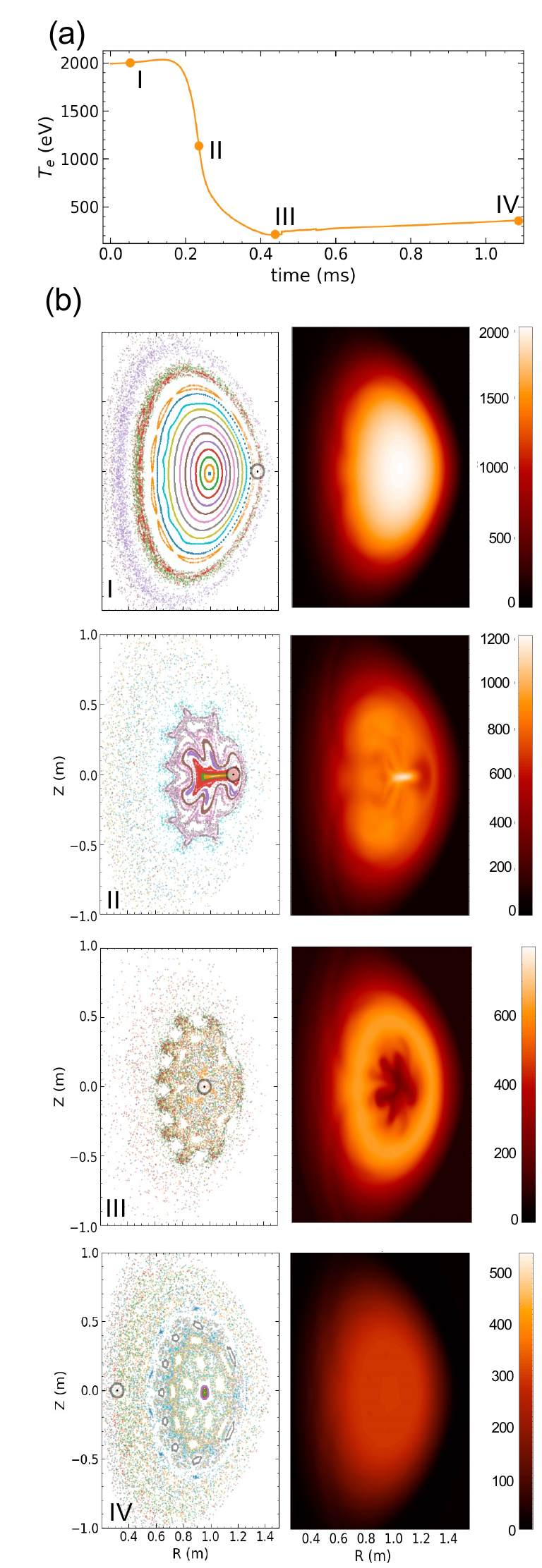}
\caption{(a) Central plasma electron temperature as a function of time and (b) 2D plasma electron temperature and Poincare plots for different times as marked in (a). Pellet velocity is $1000\,$m/s.}
\label{Fig:poincare}
\end{figure}

As can be seen from the pellet radius, all the three cases have shown a partial or incomplete pellet ablation: only $11\,\%$ of the pellet material was ablated for $v_p = 1000\,$m/s and only $21\,\%$ for $300\,$m/s. The first thing to note is that $1000\,$m/s is a high velocity for a small size device such as NSTX-U, therefore reducing the velocity from $1000\,$m/s to $300\,$ increased the ablation substantially. This also increases the radiation and produced a more significant drop in the plasma thermal energy. However, when looking at the plasma central temperature, it can be observed that it falls more abruptly than the thermal energy but, after it reaches a minimum, it increases slowly in time as the pellet finishes its path through the plasma. 

To better understand this, we show in Fig. \ref{Fig:poincare} four time slices of the plasma electron temperature together with the corresponding Poincare plot, for the case in which $v_p = 1000\,$m/s. Figure \ref{Fig:poincare}(a) shows the central temperature as a function of time, with different labels (I-IV) corresponding to $0.052,0.235,0.438,1.09\,$ms, and the time slices with 2D temperature and Poincare plots are shown in Fig. \ref{Fig:poincare}(b). Pellet position is also indicated in Poincare plots with a circle.

The plasma response and the thermal collapse due to the pellet ablated material is clearly seen in the 2D temperature plots. At panel (I), the pellet is starting to travel inside the plasma separatrix and produces field line stochastization in the outer region while the core remains unperturbed. At (II), the pellet has propagated to the $q=2.4$ surface ($r/a = 0.38$). The core electron temperature has dropped from about $2000\,$eV to about $1000\,$eV. At this point the plasma central temperature is falling very sharply.  The Poincare plot shows that the core flux surfaces are broadening and becoming partially stochastic. The field lines at the pellet position are now linked to the plasma core but not to the edge. Therefore, the heat flux that balances the pellet radiation is coming primarily from the core and, hence, the plasma temperature becomes hollow. At (III), the pellet has reached the magnetic axis. The electron temperature around the magnetic axis has dropped to about $200\,$eV, but the region surrounding the magnetic axis is at a higher electron temperature of over $500\,$eV. At this point the stochastization spreads to the edge and therefore the temperature at the center starts rising due to the hotter edge plasma. Finally, in the last time slice, panel (IV), corresponding to $t=1.09\,$ms, the pellet is almost exiting the plasma from the inboard side. The resulting plasma is starting to reform core flux surfaces but with a flattened electron temperature above $250\,$eV. 

These sequences of images show that a $2\,$mm diameter carbon pellet traveling at $1$\,km/s through a NSTX plasma with a core electron $T_e$ of $\sim 2\,$keV does not fully ablate and a full thermal quench is not attained from the injection of a single pellet of this size. This is also observed in Fig. \ref{Fig:global_vp}(c) where the thermal energy does not drop substantially to quench the plasma.

\subsection{Case 1: Scan over parallel thermal conductivity}\label{sec:scan_kappa}
\begin{figure*}[h!]
\centering
\includegraphics[width=1.9\columnwidth]{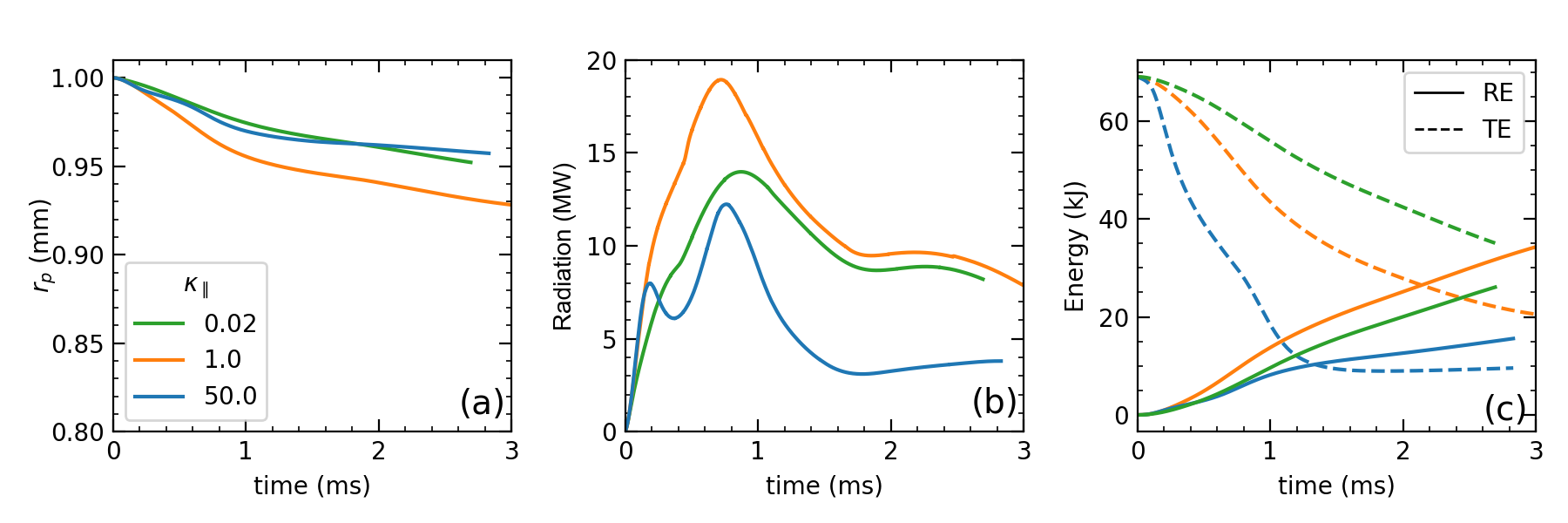}
\caption{Scanning over different parallel heat flux diffusion coefficient $\kappa_\parallel$ for a pellet velocity injected radially inward with $v_p= 300\,$m/s. (a) pellet radius, (b) radiation power and (c) radiated energy (solid) and thermal energy (dashed). }
\label{fig:global_kpar}
\end{figure*}

In the previous case the parallel thermal conductivity, $\kappa_{\parallel}$, was set to $1$ (internal units). However, the results from Fig. \ref{fig:AUG} suggest that increasing $\kappa_{\parallel}$ would increase the ablation, since the parallel heat flux will be larger. Figure \ref{fig:global_kpar} shows a scan for $\kappa_{\parallel}=50,1,$ and $0.02$. In this case we fixed the pellet velocity to $300\,$m/s.

With very low parallel heat flux $\kappa_{\parallel}=0.02$, the temperature surrounding the pellet falls due to the radiation which in turns lowers the ablation rate. Increasing the parallel thermal conductivity to $\kappa_{\parallel}=1.0$ makes the temperature surrounding pellet higher, enhancing the ablation and radiation. However, when increasing $\kappa_{\parallel}$ even more, to $\kappa_{\parallel}=50$, we observe that the ablation and radiation are reduced. 

These trends can be explained by looking at the Poincare plots which are similar to that shown in Fig. \ref{Fig:poincare}(b) for $v_p=1000\,$m/s. Before the pellet reaches the core, the stochastization links the pellet position with the plasma boundary which cools the plasma temperature faster than previous cases. Once the pellet is at the core region, the stochastization at the pellet position no longer reaches the boundary producing an enhancement in the radiation and in the ablation. However, when the pellet continues through the inner part of the plasma, the stochastization is complete and again links the pellet position with the boundary, cooling the plasma. Even though the radiated energy is smaller than in the other two cases, the thermal energy drops to a lower value. This is because a larger fraction is lost through the wall.

Contrary to the AUG-like case presented in Fig. \ref{fig:AUG}, increasing $\kappa_{\parallel}$ does not necessarily increase the ablation, as it is constrained by the stochastization that the pellet might produce.

\subsection{Case 2: Injection with a toroidal velocity}\label{sec:pellet_array}

\begin{figure}[h!]
\centering
\includegraphics[width=0.9\columnwidth]{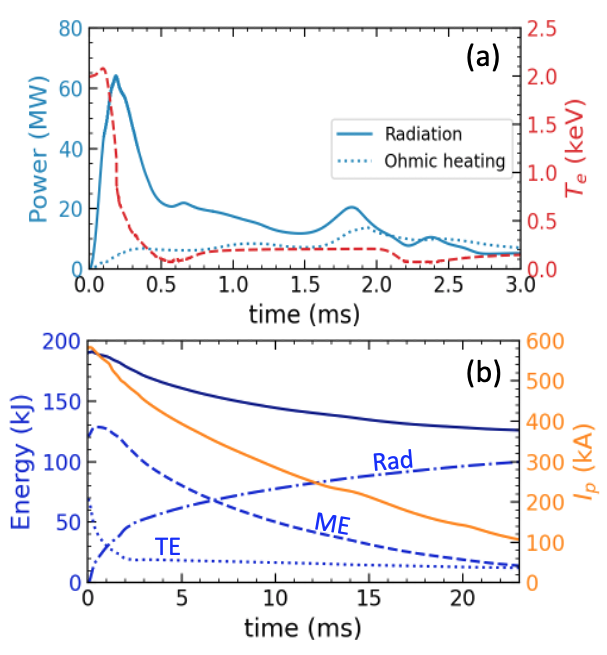}
\caption{(a) The total radiated power and the central electron temperature for the case of a tangential hollow pellet injection at $1\,$km/s, along the trajectory shown as Case 2 in Fig. \ref{Fig:top_view}. (b) Time evolution of the plasma thermal energy, TE (dotted), plasma magnetic energy, ME (dashed), and radiated energy, Rad (dash-dotted). The blue solid line is the sum of these three contributions. The plasma current is also shown in orange (solid)}
\label{fig:case2}
\end{figure}

In all the previous cases, the pellet was injected radially inward, which corresponds to Case (1) in Fig. \ref{Fig:top_view}. To increase the pellet ablation fraction, a larger radius pellet which also has a toroidal velocity component was simulated, which is labled Case (2) in Fig. \ref{Fig:top_view}. Even though \code has the capability to model the injection of several small pellets, such as would be the case with SPI, the use of a larger diameter pellet is an approximation to an array of smaller pellets since the purpose is to increase the effective surface area and therefore the ablation. In this case, we employed a $3.6\,$mm radius pellet. The pellet is hollow and the thickness was adjusted so that the amount of material is the same as in the previous cases. In this way, the surface is increased by $\sim 13$ and would be roughly similar to having an array of smaller pellets with the same total surface and material.
This case is shown in Figure \ref{fig:case2} for a pellet velocity of $1000\,$m/s and $\kappa_{\parallel}=1.0$.  

Figure \ref{fig:case2}(a) shows the radiation power and the plasma central electron temperature during the time the pellet is passing through the plasma. At $t\approx0.45\,$ms the pellet reaches the magnetic axis for the first time, showing a minimum in the radiated power as explained with the previous case. At $t\approx1.4\,$ms the pellet reaches the minimum major radius position of $0.35\,$m. Here again the radiated power presents a local minimum value since the pellet is in the inner side of the plasma. The pellet reenters the plasma from the inner side and reaches again the magnetic axis at $t \approx 2.3\,$ms and, finally, at $t \sim 2.9\,$ms the pellet hits the outer wall. The total ablated material was $\sim 32\%$, showing a significant increase from the radial injection of the smaller pellet, which was around $\sim 11\,\%$.

Figure \ref{fig:case2}(b) shows the plasma current (solid line) the plasma thermal (dotted line), magnetic (dashed line) and radiated (dash-dotted line) energy as a function of time for the entire simulation, which ran up to $23\,$ms. We can observe that the amount of ablated material $\sim 1\times10^{20}\,$atoms starts to produce a current quench (CQ), somewhat consistent with the 2D estimation. The CQ seems to be stronger than Case I in Fig. \ref{Fig:2D_global}. This is because, in 3D simulations the stochastization of the field lines enhances the heat flux loses through the wall, as explained in Sec. \ref{sec:scan_kappa}. In this case, it can be inferred to be $\sim 60\,$kJ at $t=23\,$ms (approximately $30\,\%$ of the entire initial plasma thermal and magnetic energy).

\subsection{Case 3: Shell pellet} \label{sec:shell_pellet}
The fact that in all previous simulations the ablation was incomplete, even at velocities of $\sim 300\,$m/s suggests that high velocity pellets may have the potential for penetration into the plasma core in larger machines, for example in ITER. This is the basic idea of a shell pellet, which protects the payload until it reaches the core and deposits its payload there in order to induce an inside-out thermal quench. Experiments \cite{Hollmann2019} at DIII-D have demonstrated this technique and simulations were also conducted \cite{Izzo2017} using the NIMROD code \cite{Sovinec2004}. 

As a test case to this approach, we performed a simulation which tries to resemble the injection of a hollow carbon pellet filled inside with carbon dust. The pellet was injected along the trajectory indicated with Case ``3'' in Fig. \ref{Fig:top_view}, with a velocity of $1000\,$m/s. In this case, the pellet is first ablated as usual, following the ablation formula described in Sec. \ref{Sec:pellet_modelling}, but once it reached the $R\approx 1.1\,$m position ($\sqrt{\psi_n}\approx 0.2$) at $t\sim 0.41\,$ms, we turned off the ablation, which gave values $\dot{N}\sim 10^{23}\,$atoms/sec, and switched the deposition rate to a constant (larger) value of $\dot{N}=1.14\times 10^{24}\,$atoms/sec. This constant deposition rate was continued to be centered at the (virtual) pellet position and distributed with the same Gaussian shape presented in Sec. \ref{Sec:pellet_modelling}, but the toroidal cloud size was increased from $\Delta_t = 0.5$ to $1$. This rate was maintained for $0.2755\,$ms. The total amount of Carbon deposited (including the initial ablation) was $3.8\times 10^{20}\,$atoms, which is similar to the case (II) that was presented as a proxy in Sec. \ref{sec:NSTX_2D}. After that time the deposition was turned off.

\begin{figure}[h!]
\centering
\includegraphics[width=0.95\columnwidth]{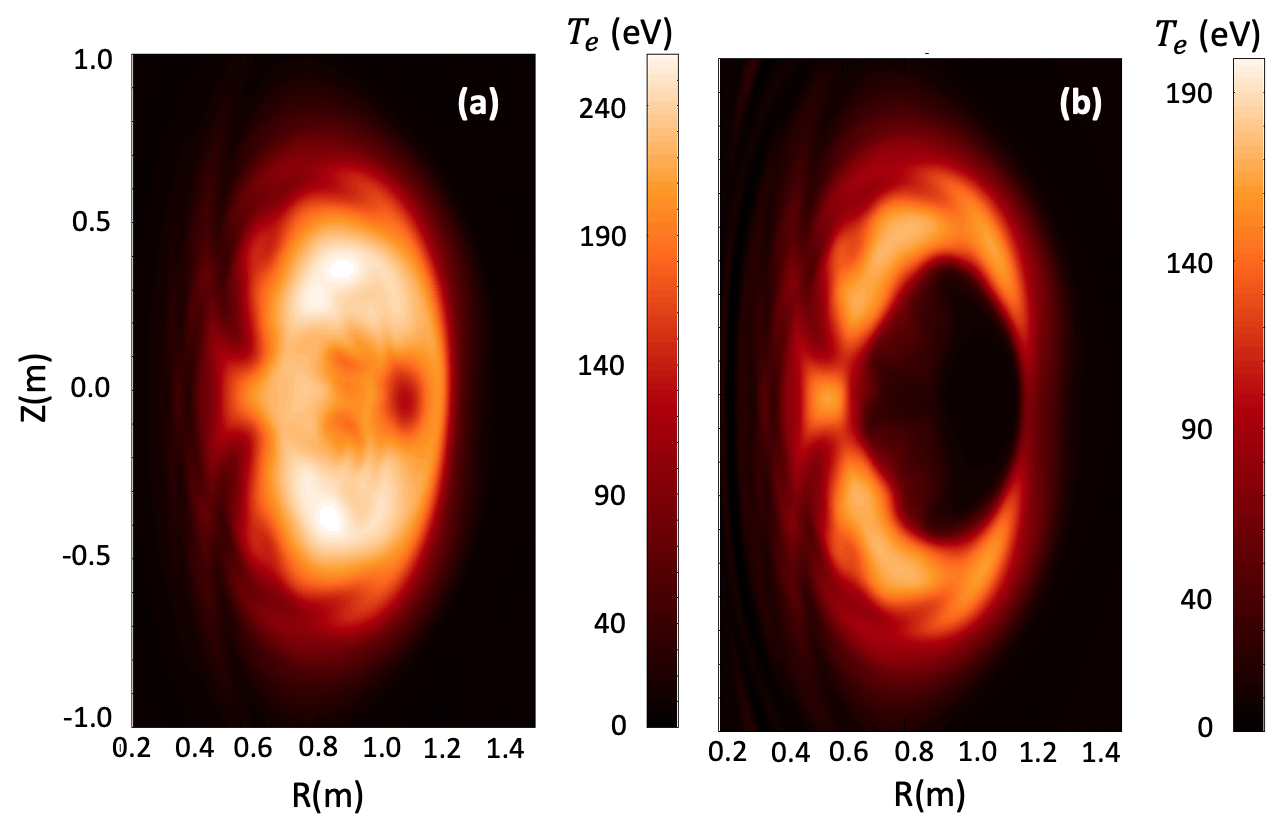}
\caption{Plasma electron temperature (a) just before turning on the large constant deposition rate, at $t=0.41\,$ms, and (b) after $0.05\,$ms of turning it on ($t=0.46\,$ms).}
\label{fig:case3_te}
\end{figure}

Figure \ref{fig:case3_te} shows the electron temperature (a) just before turning on the large deposition rate, at $t=0.41\,$ms, and (b) shortly after turning it on ($t=0.46\,$ms).
The darker region around $R\sim 1.1\,$m and $Z\sim 0$ that is observed in Fig. \ref{fig:case3_te}(a) corresponds to the pellet position and ablated cloud material that cools the surrounding plasma. It is clearly observed that after turning the large constant deposition rate on, the plasma central temperature collapses and the plasma temperature becomes hollow.

\begin{figure}[h!]
\centering
\includegraphics[width=0.9\columnwidth]{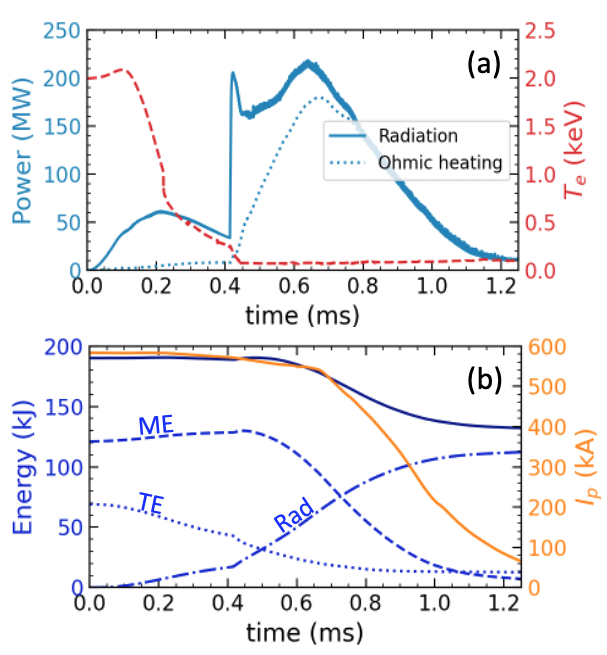}
\caption{(a) The total radiated and ohmic heating power (blue), and the central electron temperature (red) for a shell pellet injection at $1\,$km/s, following the trajectory indicated as Case 2 in Fig. \ref{Fig:top_view}. (b) Plasma thermal energy, TE, (dotted), plasma magnetic energy, ME, (dashed) and total radiated energy, Rad, (dash-dotted) are shown in blue. The solid blue line is the sum of all of them. The drop indicates the heat flux flowing into the wall. Also shown is the plasma toroidal current in a orange solid line.}
\label{fig:case3_global}
\end{figure}

Figure \ref{fig:case3_global}(a) shows, as with the previous case, the radiated power and ohmic heating (in blue) as a function of time as well as the central temperature (in red). Figure \ref{fig:case3_global}(b) shows the plasma thermal energy (TE), magnetic energy (ME) and radiated energy (Rad). The blue solid line represents the sum of the three terms. The toroidal plasma current is also shown for reference in orange.
We observe that, as in previous cases, the central temperature falls down to a couple of hundreds of eVs during the ablation phase, before the pellet reaches the core. When turning on the large constant deposition rate at $t\sim 0.41\,$ms, the central temperature collapses while the surrounding temperature remains higher and decays slower. 

Even though the amount of carbon is similar to the 2D case shown in Sec. \ref{sec:NSTX_2D}, here the plasma quench occurs much faster than in the Case II of Fig. \ref{Fig:2D_global}. One main difference here is the stochastization of the field lines, not present in 2D simulations, that enhance the heat flux towards the wall. This can be noted as the drop in the blue solid line in Fig. \ref{fig:case3_global}(b).
In the 2D case, the heat that went to the wall was almost $\sim 13\%$ of the total plasma energy, as shown in Fig. \ref{Fig:2D_global}(a), but here the total heat that reached the wall was approximately $60\,$kJ ($\sim 31\%$ of the total plasma energy content). The stochastization of the field lines, that in this case cover the entire cross section, should also prevent the formation of runaway electrons, but this analysis was not in the scope of the present paper.

\section{Summary}
\label{Sec:conclusions}

We have conducted a broad range of simulations modelling single C-pellet injections in an NSTX-U-like plasma in support of the EPI concept. This set of simulations constitute an important step towards predict-first simulations for disruption mitigation in NSTX-U and other devices, such us ITER. We have incorporated a carbon ablation model in \code and tested it in an ASDEX-U-like discharge for which data exists, obtaining very good agreement. Preliminary 2D simulations in NSTX-U suggested that the carbon content in $1\,$mm radius vitreous carbon pellet ($\sim 3.2\times 10^{20}\,$atoms) should be enough to mitigate the plasma if entirely ablated and have used this as a proxy for the amount of carbon that was considered for 3D pellet injection simulations.

We performed a wide range of 3D simulations, injecting a single C-pellet from the outer midplane, scanning over different modelling parameters such as parallel thermal conductivity and pellet velocity, among others, to show the sensitivity of the induced thermal quench and other relevant quantities. When injecting a $1$mm-radius pellet radially inward, the simulations show that the pellet is partially ablated, from $11\%$, for $v_p=1000\,$m/s, to $21\%$, for $300\,$m/s, leading to post-TQ temperature of hundreds of eVs which is not enough to produce a current quench. We also injected the pellet with an initial toroidal velocity to maximize the path length and with increased  pellet radius in order to increase the effective ablation surface. This would be similar to injecting an array of smaller pellets. In this case, in which the pellet velocity was $1000\,$m/s, the ablation increased to $\sim 32\%$ of the amount in $1$mm-pellet. Even though it was still an incomplete ablation relative to this amount, it was enough to produce a current quench with a timescale of $\sim 15-20\,$ms, showing the importance of the heat that flows into the wall due to the stochastization of the field lines.

We also have explored the injection of a shell pellet filled with carbon dust.
We have modeled the carbon dust deposition by switching the ablation rate formula to a larger constant deposition rate for a short period of time. The total amount of carbon deposited was $3.8\times 10^{20}\,$atoms. In this simulation the pellet produced a core temperature collapse followed by a fast thermal and current quench.

The broad set of scans in this work did not allow the use of very fine mesh grids and a lower density diffusion coefficient. However, future work will be focused on a particular pellet configuration and a particular discharge that includes a  $q=2$ surface. This will allow us to refine numerical modelling parameters that may affect higher toroidal modes numbers.

\ack
This work was supported by the U.S. Department of Energy under DOE Contract DE-AC02-09CH11466 and the SciDAC CTTS. This research used resources of the National Energy Research Scientific Computing (NERSC) Center, a DOE Office of Science User Facility supported by the Office of Science of the U.S. Department of Energy under Contract No. DE-AC02-05CH11231.


\section*{References}

\begin{thebibliography}{10}

\bibitem{Hender2007}
T.C~Hender \etal.
\newblock Chapter 3: {MHD} stability, operational limits and disruptions.
\newblock {\em Nuclear Fusion}, 47(6):S128--S202, 2007.

\bibitem{Commaux2016}
N.~{Commaux}, D.~{Shiraki}, L.~R. {Baylor}, E.~M. {Hollmann}, N.~W. {Eidietis},
  C.~J. {Lasnier}, R.~A. {Moyer}, T.~C. {Jernigan}, S.~J. {Meitner}, S.~K.
  {Combs}, and C.~R. {Foust}.
\newblock {First demonstration of rapid shutdown using neon shattered pellet
  injection for thermal quench mitigation on DIII-D}.
\newblock {\em Nuclear Fusion}, 56(4):046007, 2016.

\bibitem{Raman2019}
R.~Raman, W.-S. Lay, T.R. Jarboe, J.E. Menard, and M.~Ono.
\newblock {Electromagnetic particle injector for fast time response disruption
  mitigation in tokamaks}.
\newblock {\em Nuclear Fusion}, 59(1):016021, 2019.

\bibitem{Raman2021}
R.~Raman, R.~Lunsford, C.~Clauser, S.~C. Jardin, J.~E. Menard, and M.~Ono.
\newblock \textit{Prototype tests of the Electromagnetic Particle Injector-2
  for Fast Time Response Disruption Mitigation in Tokamaks}. {S}ubmitted to
  {N}ucl. {F}usion.
\newblock 2021.

\bibitem{Lyons2019}
B.~C. Lyons, C.~C. Kim, Y.~Q. Liu, N.~M. Ferraro, S.~C. Jardin, J.~McClenaghan,
  P.~B. Parks, and L.~L. Lao.
\newblock {Axisymmetric benchmarks of impurity dynamics in
  extended-magnetohydrodynamic simulations}.
\newblock {\em Plasma Physics and Controlled Fusion}, 61(6):064001, 2019.

\bibitem{Ferraro2019}
N.M. Ferraro, B.C. Lyons, C.C. Kim, Y.Q. Liu, and S.C. Jardin.
\newblock {3D two-temperature magnetohydrodynamic modeling of fast thermal
  quenches due to injected impurities in tokamaks}.
\newblock {\em Nuclear Fusion}, 59(1):016001, 2019.

\bibitem{Sergeev2006}
V.~Yu. Sergeev, O.~A. Bakhareva, B.~V. Kuteev, and M.~Tendler.
\newblock {Studies of the impurity pellet ablation in the high-temperature
  plasma of magnetic confinement devices}.
\newblock {\em Plasma Physics Reports}, 32(5):363--377, 2006.

\bibitem{Jardin2012}
S.~C. Jardin, N.~Ferraro, J.~Breslau, and J.~Chen.
\newblock {Multiple timescale calculations of sawteeth and other global
  macroscopic dynamics of tokamak plasmas}.
\newblock {\em Computational Science {\&} Discovery}, 5(1):014002, 2012.

\bibitem{Parks1977}
P.~B. Parks, R.~J. Turnbull, and C.~A. Foster.
\newblock {A model for the ablation rate of a solid hydrogen pellet in a
  plasma}.
\newblock {\em Nuclear Fusion}, 17, 1977.

\bibitem{Parks1978}
P.~B. Parks and R.~J. Turnbull.
\newblock Effect of transonic flow in the ablation cloud on the lifetime of a
  solid hydrogen pellet in a plasma.
\newblock {\em The Physics of Fluids}, 21(10):1735--1741, 1978.

\bibitem{Kuteev1984}
B.~V. Kuteev, V.~Y. Sergeev, and L.~D. Tsendin.
\newblock {Interaction of Carbon pellets with a hot plasma}.
\newblock {\em Sov. J. Plasma Phys.}, 10(6):675, 1984.

\bibitem{Lunsford2019}
R.~Lunsford, A.~Bortolon, R.~Maingi, D.K. Mansfield, A.~Nagy, G.L. Jackson, and
  T.~Osborne.
\newblock Supplemental elm control in iter through beryllium granule injection.
\newblock {\em Nuclear Materials and Energy}, 19:34--41, 2019.

\bibitem{Sergeev1994}
V.~Y. Sergeev, S.~M. Egorov, B.~V. Kuteev, and I.~V. Miroshnikov.
\newblock {Plasma Diagnostics on ASDEX Upgrade by means of Carbon Pellet
  injection}.
\newblock {\em ECA: 21st European Conf. on Controlled Fusion and Plasma
  Physics}, 18B:1364, 1994.

\bibitem{Hollmann2019}
E.~M. Hollmann, P.~B. Parks, D.~Shiraki, N.~Alexander, N.~W. Eidietis, C.~J.
  Lasnier, and R.~A. Moyer.
\newblock Demonstration of tokamak discharge shutdown with shell pellet payload
  impurity dispersal.
\newblock {\em Phys. Rev. Lett.}, 122:065001, 2019.

\bibitem{Izzo2017}
V.~A. Izzo and P.~B. Parks.
\newblock Modeling of rapid shutdown in the diii-d tokamak by core deposition
  of high-z material.
\newblock {\em Physics of Plasmas}, 24(6):060705, 2017.

\bibitem{Sovinec2004}
C.R. Sovinec, A.H. Glasser, T.A. Gianakon, D.C. Barnes, R.A. Nebel, S.E.
  Kruger, D.D. Schnack, S.J. Plimpton, A.~Tarditi, and M.S. Chu.
\newblock {Nonlinear magnetohydrodynamics simulation using high-order finite
  elements}.
\newblock {\em Journal of Computational Physics}, 195(1):355--386, 2004.

\end{thebibliography}

\end{document}